# Influence of Demagnetization Effect on Giant Magneto Impedance

S. K. Ghatak
Department of Physics & Meteorology
Indian Institute of Technology, Kharagpur 721302, India

Abstract : The large change in electromagnetic impedance in ferromagnetic state of soft magnetic metals in presence of biasing magnetic field is associated with change in screening of electromagnetic field. The screening depends on the permeability of the metal. Apart from dependence on intrinsic properties of metal the permeability depends on size of the sample. It is observed that the decrease in MI in amorphous ferromagnetic ribbon of $Fe_{40}Ni_{40}B_{20}$ alloy is large for long sample whereas corresponding change is small for short one with same biasing field. As intrinsic magnetic properties and bias field are same and the demagnetization factor increases with reduction of length of the sample the reduction of MI effect is associated with demagnetization field.

**Introduction:**

Electromagnetic response of metal in ferromagnetic state depends sensitively of magnetic softness ,and a large change in impedance in presence of external magnetic field (referred to giant magnetic impedance (GMI)) has been observed in ferromagnetic metallic alloy of different shapes e.g. wire [1,2] , ribbon [3-6] ,film [7]. The external static magnetic field induces a change in complex permeability which has strong influence of current distribution inside the sample. For soft ferromagnetic material this change in permeability can be very large as small field alters internal field in significant manner and thereby causing large change in current distribution which in turn produces a large change in impedance. So, the magnetic impedance (MI) of ferromagnetic sample would depend on parameters - intrinsic or extrinsic in character that determine permeability. Large value of permeability is often found in transition metal-metalloid metallic glass with low hystersis field and the GMI effect in presence of few Oe bias field has been observed in Fe-based ferromagnetic metallic ribbon [8-11]. Recent results of GMI phenomenon in soft ferromagnetic metal [12-13].The demagnetizing field that depends on shape and size of sample affect the magnetic response in ferromagnetic state and therefore, the measured permeability is often smaller than the actual one. The impedance of ferromagnetic ribbon is measured using either direct or indirect method. In direct method, a.c excitation current flows along ribbon length and the impedance is obtained from measured potential drop across voltage probe. In this case, the a.c magnetic field is in direction of width of the ribbon and the magnetic response that determines screening of electric field is influenced by demagnetization effect of ribbon. In indirect or non-contact method the a.c magnetic field is created by a.c. current flowing through a signal coil wrapped over the sample and the impedance of the sample is obtained from that of coil. In this situation the direction of the magnetic and electric field is interchanged compared to that in earlier case. To quantify the influence of size of the sample on MI, the MI of amorphous ferromagnetic ribbon of nominal composition $Fe_{40}Ni_{40}B_{20}$ is measured varying length of the ribbon. It is found that the relative change in MI is much reduced when the sample size is decreased so that the demagnetization factor plays prominent role in determining magnetic response of the material. A simple model for MI is presented that qualitatively support the experimental observation.

**Experimental:**

The amorphous ribbon (Lx 3mm x120μm) of nominal composition $Fe_{40}Ni_{40}B_{20}$ was used. A small coil of rectangular geometry (50 turn ??) is wrapped over the ribbon of length L ,larger than coil length. This inductor together with a standard series resistor is excited by small a.c. current derived from constant current source .The excitation magnetic field is generated by through coil. The real and imaginary components of impedance, Z = R + jX, where reactance X = ωl (l is the inductance ), is derived from the measurement of potential drop and current through coil using Lock-in Amplifier ( Model -7280 DSP-LIA –signal recovery).The d.c biasing field is produced by Helmholz coil and the sample is placed within uniform region of field with length of the ribbon is along field. The assembly is oriented in such a way that the Earth's magnetic field is perpendicular to d.c field. All the measurements were performed at room temperature The resistance $R_s$, reactance $X_c$ and impedance $Z_c$ of the sample presented here are obtained after eliminating the corresponding values of empty coil. The amplitude of a.c. current is kept around 4mA and the response of the sample is found to be linear around this current.

**Results & Discussions:**

The resistance R , reactance X and impedance Z of the $Fe_{40}Ni_{40}B_{20}$ ribbon at frequency 1MHz are plotted in Fig.1 as function of d.c. biasing field H for length 60mm and 5.5mm. At zero field R< X and hence Z is slightly higher compared to X for both samples and the ratio of resistances is much higher than that of corresponding lengths of the ribbon. Moreover, the field behaviour of R is different for two samples. For longest sample, the resistance R decreases from ~ 53Ω at zero field to ~ 3Ω with small field of few Oe. In contrast the resistance variation within same field is only few percent for smallest sample. Similar nature of variation of reactance X (impedance Z) has been seen except that higher field is needed for same extent of reduction. In fig.2 the results of relative change of impedance (dZ/Z(0)) at frequency 1MHz for samples of different length is shown. Except of length all sample are identical. At low field region dZ/Z(0) decreases slowly for sample of smallest length and sharpness of fall goes up with increase in length of the sample. At higher field the (dZ/Z(0)) tends to saturate at value that depends on sample length. For long samples nearly complete reduction of Z is found. The field dependence of MI in ferromagnetic state of metal originates from the screening of electromagnetic field by response of magnetic state. The response is quantified by the permeability of material. The screening is determined by the skin depth $\delta_m = \delta_0/\sqrt{\mu}$ where the skin depth of non-magnetic metal $\delta_0 = (2/\omega\mu_0\,\sigma)^{1/2}$ at frequency $\omega$ , and $\mu_0$ , $\sigma$ being permeability of vacuum and d.c. conductivity of the material, respectively.

The permeability $\mu$ depends on number of intrinsic and extrinsic parameters. For soft ferromagnetic material like transition metal-metalloid metallic glass the permeability is very large in absence of any biasing field that hindered the magnetization process and this leads smaller skin depth and larger impedance. In presence of biasing d.c. field larger than anisotropy field the magnetization of the sample orients parallel to field and responds feebly to a.c. field resulting small value of permeability. In this situation impedance falls as field penetrates whole sample. The magnetic susceptibility $\chi$ ( $\mu$ -1) of finite size ferromagnetic sample depends on its shape and size and such dependence is expressed in terms of demagnetization factor N as $\chi_a^{-1} = \chi_i^{-1} + N$ . For long sample with N~0, the measured susceptibility $\chi_a$ is close to real one $\chi_i$ whereas it is demagnetization dominated and smaller for sample of smaller dimension.

The above notion is further supported by a model calculation of MI developed earlier in relation to stress-induced impedance [14]. The magnetic susceptibility of soft ferromagnetic substance is a complex function of frequency, anisotropy field, saturation magnetization ,dimension of sample. The complex susceptibility and hence the impedance can be derived from dynamics of magnetization, and the Maxwell's equations. The Landau-Lifsitz-Gilbert equation is customarily used to describe the macroscopic dynamics of magnetization M in ferromagnet [13-15 ]

$$\dot{\boldsymbol{M}} = \gamma\left(\boldsymbol{M} \times \boldsymbol{H}_{eff}\right) - \frac{\alpha}{M_s}\left(\boldsymbol{M} \times \dot{\boldsymbol{M}}\right) - \frac{1}{\tau}\left(\boldsymbol{M} - \boldsymbol{M}_0\right) \qquad (1)$$

where γ is gyromagnetic ratio, α coefficient of Gilbert damping and τ is the relaxation time determining the Bloch-Bloembergen damping. In contrast to the Gilbert damping this damping does not preserve the magnitude of M. This process is more relevant for amorphous materials with imperfect ferromagnetic order. The effective magnetic field $\boldsymbol{H}_{\text{eff}}$ is sum of external (d.c and a.c) anisotropy and demagnetization fields. In presence of small a.c. magnetic field along the ribbon length the complex permeability χ =(m/h) where m is induced maganetization due to a.c. field of magnitude h. The induced magnetization $\vec{m}$ is obtained from the equation

$$i\omega^{*}\vec{m} + (i\alpha\omega\frac{\vec{M}_0}{\vec{M}_s} + \gamma\vec{H}_{eff\,0}) \times \vec{m} = \gamma(\vec{M}_0 \times \vec{h}_{eff}) \qquad (2)$$

where $\omega^{*} = \omega - i/\tau$ . Here $\vec{H}_{eff\,0}$ is value of effective field in absence of excitation. The effective excitation field $\vec{h}_{eff} = \vec{h} - N\vec{m} + \vec{h}_{an}$ where N is demagnetization factor and $\vec{h}_{an} = H_{an}\vec{e}(\vec{e}.\vec{m})$ , $\vec{e}$ ,unit vector defining anisotropy field. The equilibrium magnetization $\vec{M}_0$ is given by the condition that total torque

$$\vec{M}_0 \times \vec{H}_{eff\,0} = 0 \qquad (3).$$

These equations are solved considering uniaxial anisotropy, magnetization and external fields are in the plane of the ribbon and the z-axis is taken along length of the ribbon (Fig.5).Then the complex susceptibility is given by

$$\chi = \frac{H_{ey} M_s \sin\theta}{\left( -\frac{\omega^{*2}}{\gamma^2} - A_y H_{ez} + H_{ey} A_z \right)} \qquad (4)$$

where

$$A_z = H_{ey} + \beta_z M_s \cos\theta + (N - \alpha_z) M_s \sin\theta$$
$$A_y = -H_{ez} - \alpha_y M_s \sin\theta + \beta_y M_s \cos\theta$$

$$H_{ey} = -\frac{-i\omega\alpha \sin(\theta)}{\gamma} - H_a \sin(\theta_a)\cos(\theta_a - \theta)$$

$$H_{ez} = -\frac{i\omega\alpha \cos(\theta)}{\gamma} + H\cos(\theta) + H_a \cos(\theta_a)\cos(\theta_a - \theta)$$

$$\alpha_y = -\frac{H_a}{M_s}\sin(\theta_a)\cos(\theta_a) \qquad \alpha_z = \frac{H_a \cos^2\theta_a + H}{M_s}$$

$$\beta_y = \frac{H_a}{M_s}\sin^2(\theta_a) + \frac{H}{M_s} \qquad \beta_z = \frac{-H_a \cos(\theta_a)\sin(\theta_a)}{M_s}$$

The impedance of ribbon excited by a.c. magnetic field of magnitude *h* along the length of the ribbon can be obtained from the measured potential drop $V = -i\omega\int \mu h dS$ across the coil wrapped over the ribbon. The complex permeability $\mu = 1 + \chi$ is given by eq. 4. Here *dS* is elemental surface area parallel to induced magnetic induction assumed to be linear in field *h*. From the Maxwell's equation for *h* the impedance $Z = V/I$ of the ribbon is found to be [ 14]

$$Z = -i\omega L_0 (\mu \frac{\tanh(kd)}{kd} - 1) \qquad (5)$$

where the wave-vector $k = (1+i)(\mu_0 \mu \omega \sigma / 2)^{1/2}$, σ being d.c. conductivity.

Based on the above theoretical model and using typical values of parameters $\theta_a = 45^0$, $\alpha = 50$, $M_s = 6.5\text{x}10^5$ A/m, $\tau = 10^{-5}$ s, $\sigma = 10^5$ Ω-m and $\omega_0 = 10^6$ rad/s and thickness of the ribbon = 15 μm the susceptibility and impedance have been obtained as function of biasing field H for typical small value of $H_a$ =50A/m. As the impedance is determined by the screening which in turn associated with the magnetic response the susceptibility is pertinent parameter for the magneto-impedance effect. The magnitude of complex susceptibility χ varies strongly (Fig. 3 ) in presence of small field for long sample with near zero demagnetization factor N and tends to saturate at small value at higher field ( larger than $H_a$). With increase in N the susceptibility drops to lower value and approaches to the demagnetization dominated value $N^{-1}$ value. It also becomes less susceptible to field H. The relative change $dZ/Z_0$ defined by $(Z_H - Z_0)/Z_0$ as obtained from the calculation is depicted in Fig.4 for same values of N $=10^{-5}, 10^{-3}, 5\text{x}\ 10^{-3}, 10^{-2}, 2\text{x}10^{-2}$. For near zero value of N the decrease in impedance nearly 98% for field H which is six times the anisotropy field $H_a$. The slope of $dZ/Z_0$ at small field is large and negative. As N increases the magnitude of the slope falls in this region and extent of decrease in Z also comes down. In contrast to saturation at large H ,$dZ/Z_0$ exhibits decreasing tendency. For N= $2\text{x}10^{-2}$,$dZ/Z_0$ is around 63% at H=300A/m. These scenario can be understood from the consideration of screening and field dependence of skin depth in metal. At zero bias field and for N ~0, the induced magnetization due to a.c.. field along ribbon length is large due to magnetic softness of the sample. This large response leads to smaller value of skin depth which in turn reduces effective current carrying area of the sample and hence results a large impedance. In presence of large bias field larger than anisotropy field the magnetization orients nearly parallel to ribbon length and thereby very small a.c. induction is produced by small exciting field along same direction. So in this situation the response (susceptibility) is much reduced and the current flows through entire cross-section of the sample due to larger skin depth and hence impedance is nearly same as that of non-magnetic metal and is small. For finite value of N, the measured response in ferromagnetic state depends on N and is reduced. Thus the demagnetization effect caused by the sample geometry constraint lowers the screening efficiency of ferromagnetic metal, and in such case the observed magneto-impedance effect becomes smaller. These results are in tune with the observed one and points out the importance of role of sample geometry in magneto –impedance effect. Note that smaller value of GMI ($dZ/Z_0$) often observed in direct method of measurement is perhaps due to demagnetization factor as magnetic field is usually along smaller dimension of the sample.

Conclusion: The influence of the geometry of the sample on the magneto-impedance effect in soft ferromagnetic metal has been demonstrated. The demagnetization field reduces field sensitivity and the GMI factor. The model calculation of MI corroborates the experimental observation.

Acknowledgement: The financial support from CSIR, Government of India and technical help from S.Ghosh are gratefully acknowledged.

# Figures

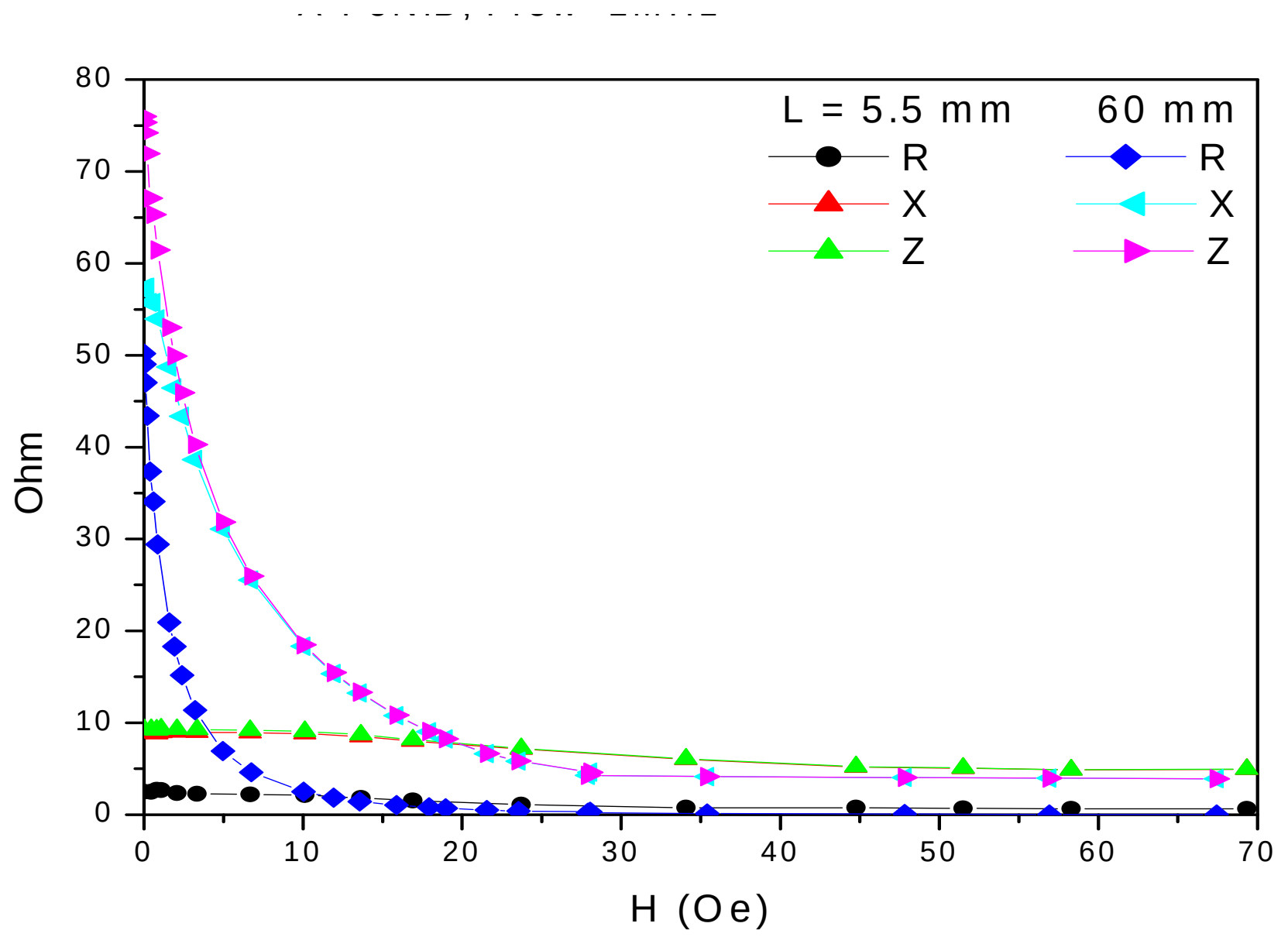

Fig.1 Plot of field variation of resistance R, Reactance X, and Impedance Z of a-$Fe_{40}Ni_{40}B_{20}$ ribbon of length L=5.5 and 60 mm at frequency 1 MHz

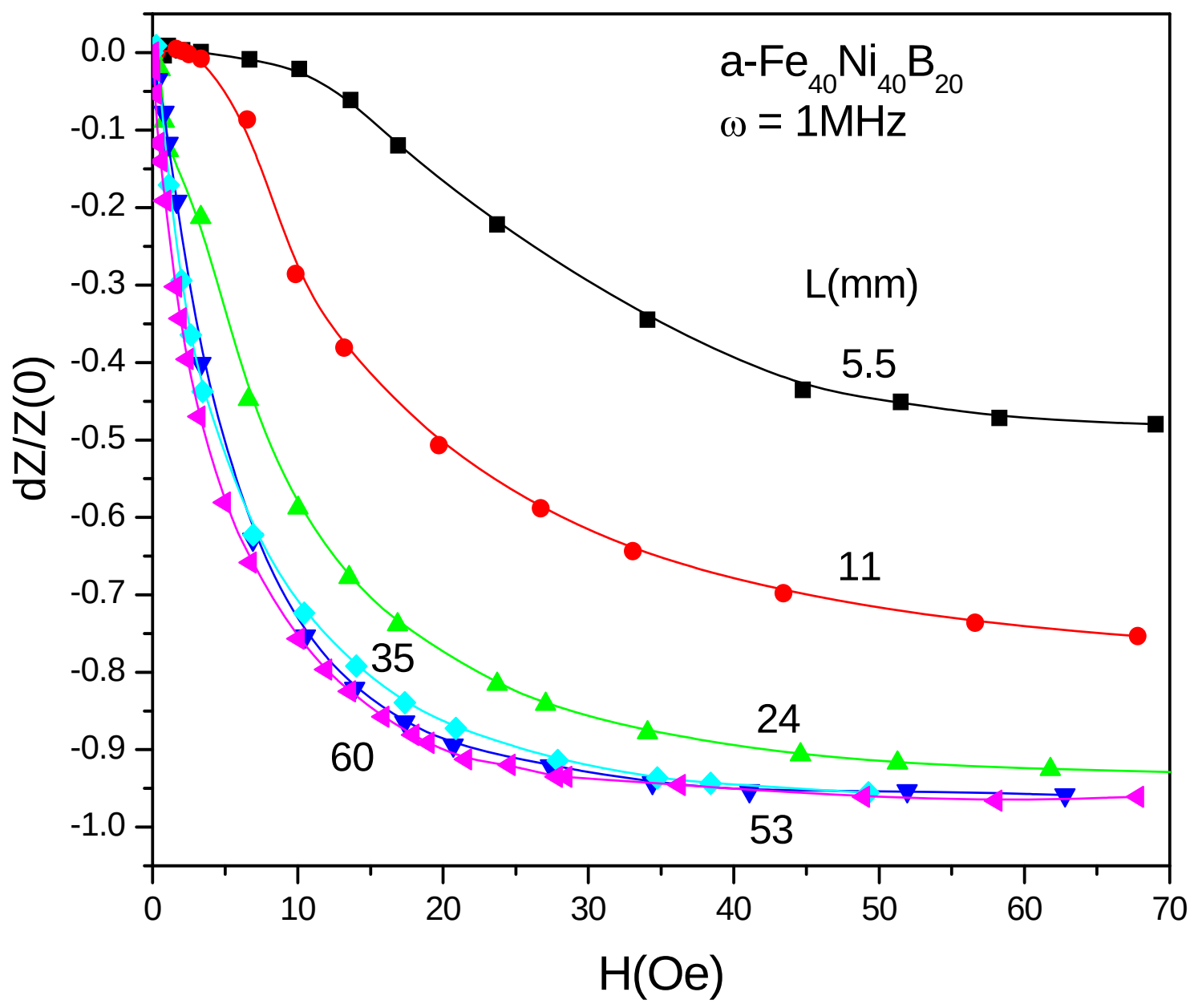

Fig.2 Relative change dZ/Z(0) of impedance of a-$Fe_{40}Ni_{40}B_{20}$ ribbon as a function of d.c magnetic field H for different length of ribbon. The number near to each curve indicates length L of ribbon in mm.

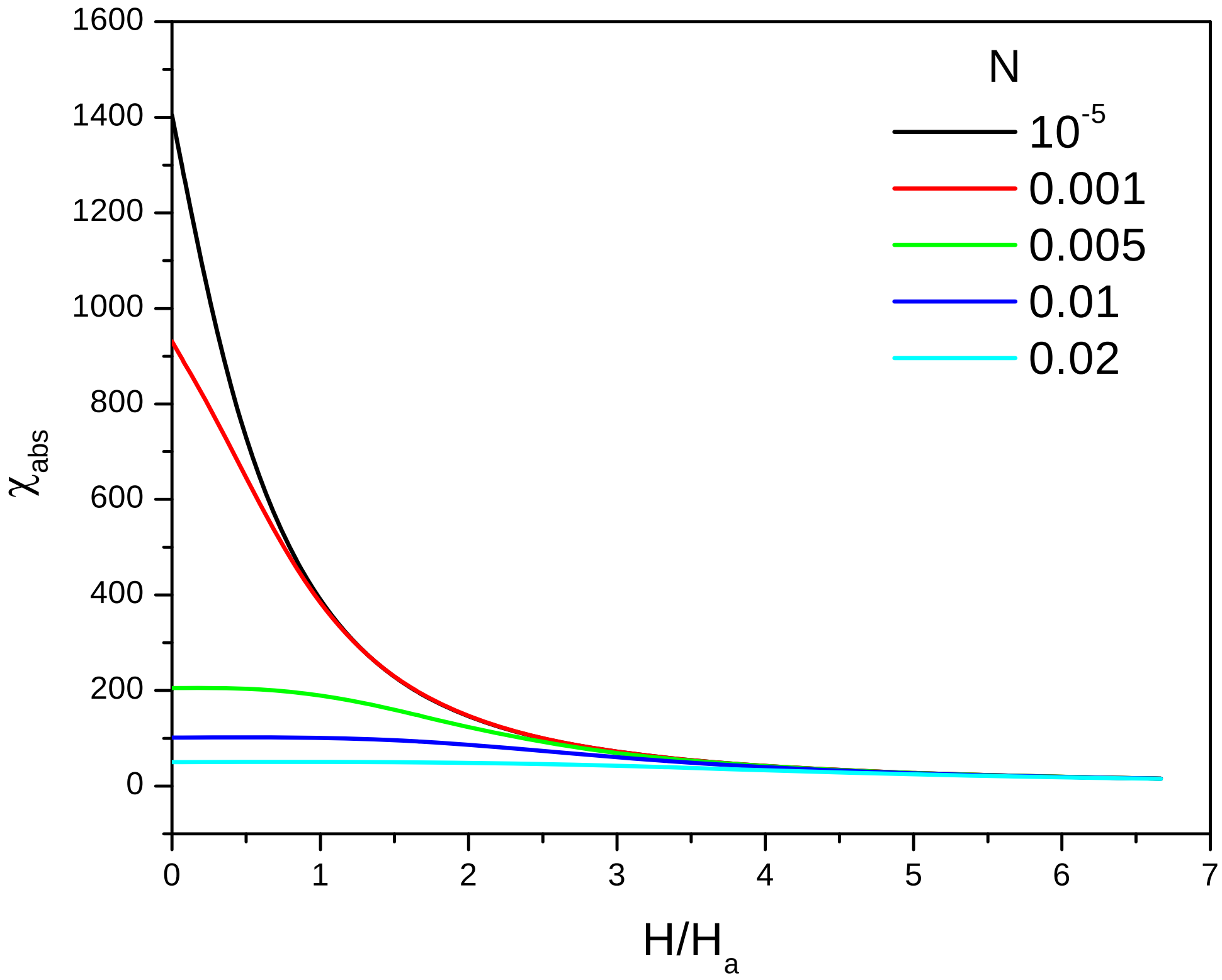

Fig.3: Variation of absolute value of magnetic susceptibility $\chi_{abs}$ with reduced field $H/H_a$ for different values of demagnetization factor N.

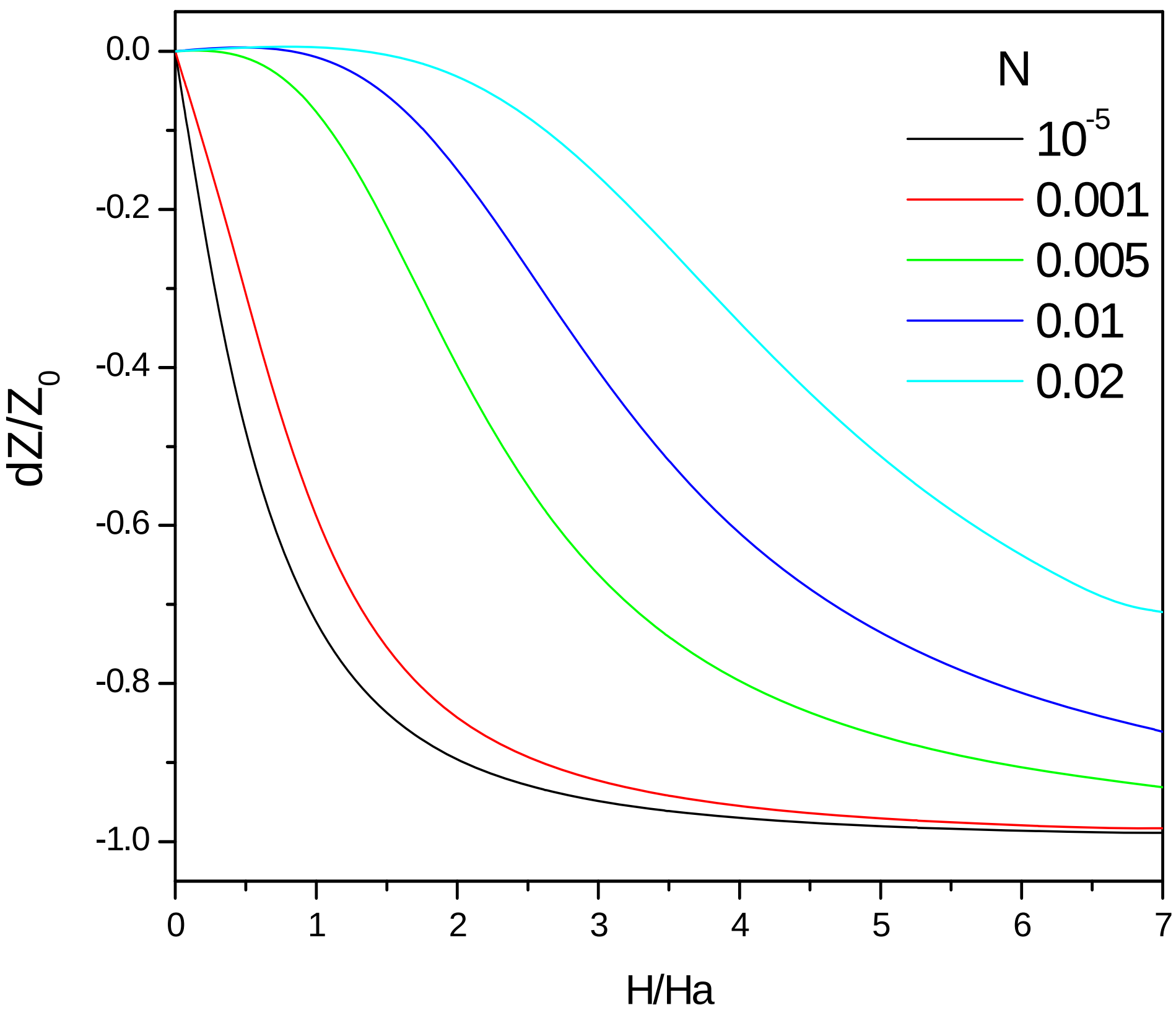

Fig.4: Relative decrease ($dZ/Z_0 = [Z_H - Z_0]/ Z_0$ ) in impedance with reduced field $H/H_a$ for different values of demagnetization factor N .